\documentclass[prl,showpacs,twocolumn,floats,10pt,aps,citeautoscript,longbibliography,superscriptaddress]{revtex4-2}
\usepackage{microtype} 
\usepackage{graphicx}
\usepackage{enumerate}
\usepackage{bm}
\usepackage{amsmath}
\usepackage{amssymb}
\usepackage{color}
\usepackage{upgreek}
\usepackage[normalem]{ulem}
\usepackage{bbm}  
\usepackage{xspace} 
\usepackage{array}
\usepackage{multirow}

\allowdisplaybreaks 

\newcommand{\mb}[1]{\ensuremath{\mathbf{#1}}}

\newcommand{\mr}[1]{\ensuremath{\mathrm{#1}}}

\newcommand{\la}{\ensuremath{\langle}}
\newcommand{\ra}{\ensuremath{\rangle}}

\newcommand{\ket}[1]{\ensuremath{| #1 \rangle}}

\newcommand{\ua}{\ensuremath{{\uparrow}}}
\newcommand{\da}{\ensuremath{{\downarrow}}}

\newcommand{\mi}{\mathrm{i}} 

\newcommand{\Gf}{\ensuremath{G^f}}

\newcommand{\Af}{\ensuremath{A^f}}

\newcommand{\OSMp}{\ensuremath{\text{OSM}^\prime}\xspace}
\newcommand{\EPC}{\ensuremath{\mr{\scriptscriptstyle{EPC}}}}
\newcommand{\VTHF}{\ensuremath{\mr{\scriptscriptstyle{VTHF}}}}
\newcommand{\nonloc}{\ensuremath{\mr{\scriptstyle{nonloc}}}}

\newcommand{\MBZ}{\ensuremath{\mr{\scriptscriptstyle{1MBZ}}}}
\newcommand{\THF}{\ensuremath{\mr{\scriptscriptstyle{THF}}}}
\newcommand{\meV}{\ensuremath{\text{meV}}\xspace}
\newcommand{\RG}
{\ensuremath{\mr{\scriptscriptstyle{RG}}}}
\newcommand{\tot}{\ensuremath{\mr{tot}}}
\newcommand{\CNP}{\ensuremath{\mr{CNP}}}

\newcommand{\Eq}[1]{Eq.~\eqref{#1}}

\newcommand{\Fig}[1]{Fig.~\ref{#1}}

\RequirePackage[
  hyperindex,colorlinks,bookmarksnumbered,
  plainpages=true,pdfstartview=FitH,hypertexnames=false]{hyperref}
\hypersetup{linkcolor=blue,urlcolor=blue,citecolor=blue} 
\usepackage[hypertexnames=false]{hyperref}
\usepackage[all]{hypcap}

\makeatletter
\def\maketitle{
\@author@finish
\title@column\titleblock@produce
\suppressfloats[t]}
\makeatother

\begin{document} 

\title{
Hundness in twisted bilayer graphene: correlated gaps and pairing
}

\author{Seongyeon Youn}
\thanks{These authors contributed equally to this work.}
\affiliation{Department of Physics and Astronomy, Seoul National University, Seoul 08826, Korea}
\affiliation{Center for Theoretical Physics, Seoul National University, Seoul 08826, Korea}

\author{Beomjoon Goh}
\thanks{These authors contributed equally to this work.}
\affiliation{Department of Physics and Astronomy, Seoul National University, Seoul 08826, Korea}
\affiliation{Center for Theoretical Physics, Seoul National University, Seoul 08826, Korea}
\affiliation{Institute for Data Innovation in Science, Seoul National University, Seoul 08826, Korea}

\author{Geng-Dong Zhou}
\affiliation{International Center for Quantum Materials, School of Physics, Peking University, Beijing 100871, China}

\author{Zhi-Da Song}
\email{songzd@pku.edu.cn}
\affiliation{International Center for Quantum Materials, School of Physics, Peking University, Beijing 100871, China}
\affiliation{Hefei National Laboratory, Hefei 230088, China}
\affiliation{Collaborative Innovation Center of Quantum Matter, Beijing 100871, China}

\author{Seung-Sup B.~Lee}
\email{sslee@snu.ac.kr}
\affiliation{Department of Physics and Astronomy, Seoul National University, Seoul 08826, Korea}
\affiliation{Center for Theoretical Physics, Seoul National University, Seoul 08826, Korea}
\affiliation{Institute for Data Innovation in Science, Seoul National University, Seoul 08826, Korea}

\date{\today}

\begin{abstract}
We characterize gap-opening mechanisms in the topological heavy fermion (THF) model of magic-angle twisted bilayer graphene (MATBG), with and without electron-phonon coupling, using dynamical mean-field theory (DMFT) with the numerical renormalization group (NRG) impurity solver.
In the presence of symmetry breaking associated with valley-orbital ordering (time-reversal-symmetric or Kramers intervalley coherent, or valley polarized), spin anti-Hund and orbital-angular-momentum Hund couplings, induced by the dynamical Jahn--Teller effect, result in a robust pseudogap at filling $2 \lesssim |\nu| \lesssim 2.5$. We also find that Hundness enhances the pairing susceptibilities for $1.6 \lesssim |\nu| \lesssim 2.8$, which might be a precursor to the superconducting phases neighboring $|\nu| = 2$.

\medskip

\noindent
DOI:

\medskip

\end{abstract}

\maketitle

\textit{Introduction.---}%
Magic-angle twisted bilayer graphene (MATBG)~\cite{Bistritzer2011} hosts an interplay of band topology~\cite{zou_band_2018,Ahn2019, Po2019, Song2019,liu_pseudo-landau-level_2019, Song2021}, strong electronic correlations~\cite{Huang2019, Soejima2020, Chen2021, Liao2021, Parker2021, Hofmann2022, Wang2023, Hu2023:DMFT, Rai2024, Zhou2024,  Datta2023, Xiao2024}, and electron-phonon coupling (EPC)~\cite{Wu2018,Lian2019,Blason2022,Liu2024,Dodaro2018,Angeli2019,Wang2024:Molecular,Shi2024,Wang2024,Kwan2023,angeli_jahnteller_2020,chen_strong_2023}.
Many experiments observed correlated insulators at filling $\nu = \pm 2$ per moir\'{e} unit cell (MUC) with respect to the charge neutrality point (CNP)~\cite{Cao2018correlated, Xie2019:STM, Choi2019:STM, Liu2021:Science,Saito2020,stepanov_untying_2020, Lu2019,Cao2020, Oh2021,  Nuckolls2023,Choi2021, Kerelsky2019, cao_nematicity_2021}
and unconventional superconductivity near them~\cite{Cao2018unconventional,Cao2020, stepanov_untying_2020, Oh2021, Nuckolls2023,cao_nematicity_2021,Lu2019, tanaka_superfluid_2024}.
Meanwhile, the insulating behavior is weaker or invisible at $|\nu|=1, 3$~\cite{Cao2018correlated,Xie2019:STM, Lu2019,Choi2019:STM,stepanov_untying_2020, Saito2020,Cao2020,Liu2021:Science, Oh2021, Nuckolls2023}.
Scanning tunneling microscopy (STM) studies of the insulator at $|\nu| = 2$ revealed a pseudogap~\cite{Oh2021, Choi2021, Nuckolls2023} accompanied by side peaks at its edges~\cite{Oh2021} and observed a time-reversal-symmetric intervalley coherent (TIVC) order~\cite{Nuckolls2023} in ultralow-strain samples, which is not energetically favored over others, such as valley polarized (VP) or Kramers intervalley coherent (KIVC) orders, without considering EPC~\cite{Shi2024,Wang2024,Kwan2023,Po2018, Kang2019, Xie2020, Zhang2020, Bultink2020, Liu2021, Lian2021}.

Recent dynamical mean-field theory (DMFT) calculations~\cite{Hu2023:DMFT, Datta2023, Rai2024, Zhou2024} elucidated key features of MATBG by capturing strong correlations: the formation and screening of local moments, cascade transitions, and their temperature dependence, to name a few.
However, they found a gap at $|\nu| = 1$ that is more significant than $|\nu| = 2$, which implies extra ingredients are necessary.
Also, it has been unclear how Mott-like features~\cite{Cao2018correlated,Kerelsky2019} arise on top of the observed TIVC order, given that symmetry breaking often competes with dynamical correlations.

In this Letter, we demonstrate that the interplay between EPC-induced (anti-)Hund couplings and valley-orbital ordering is responsible for opening a pseudogap with side peaks, over an extended region of $2 \lesssim |\nu| \lesssim 2.5$.
We incorporate the latter by downfolding the topological heavy fermion (THF) model~\cite{Song2022, Shi2022, Arbeitman2024, Singh2024, Lau2023, Chou2023:PRB, Chou2023:PRL, Hu2023, Merino2024, Calugaru2024, Hu2024, Hu2023:DMFT, Zhou2024, Rai2024} of MATBG, imposing a valley-orbital order.
By solving the downfolded model using DMFT~\cite{Georges1996} with the numerical renormalization group (NRG)~\cite{Wilson1975,Bulla2008,Weichselbaum2007,Weichselbaum2012:NRG,Lee2016:MuNRG1,Lee2017:MuNRG2,Kugler2022} impurity solver, we reveal that the effective Hund couplings invoke the Janus effect~\cite{Werner2009,deMedici2011:PRB,deMedici2011:PRL,Georges2013,Stadler2019} that favors gap opening at the half filling of the active interacting orbitals, corresponding to the range of $|\nu|$.
Hundness also facilitate fluctuations of electron pairs with a specific symmetry, over a range of $\nu$ overlapping with the reported superconducting domes~\cite{Cao2018unconventional, Cao2020, Oh2021, cao_nematicity_2021,Nuckolls2023,tanaka_superfluid_2024}.

\textit{Valley-orbital-ordered THF model.---}%
The THF model~\cite{Song2022,Shi2022} consists of two interacting $f$ orbitals ($\alpha = 1,2$) localized at the AA-stacking region of each MUC, hybridized with four topological Dirac $c$ bands, per spin ($\sigma = \ua,\da$) and valley ($\eta = \pm$).
At CNP ($\nu = 0$), the Kondo screening of $f$ electrons is suppressed since the density of states for the $c$ bands vanishes at the Fermi level. 
Thus $f$ electrons are susceptible to valley-orbital ordering via Ruderman--Kittel--Kasuya--Yosida (RKKY) interactions~\cite{Song2022, Bultink2020, Bernevig2021:TBG5, Hofmann2022, Zhou2024}.
When doped to $\nu > 0$, excess electrons experience Kondo correlations due to finite hybridization.
To incorporate both effects, we first downfold the THF model onto the active subspace of excess electrons, on top of the valley-orbital order at CNP chosen among the VP, KIVC, and TIVC states.

The downfolded, valley-orbital-ordered THF (VTHF) model for $\nu > 0$~\cite{Zhou2024, Youn2025} has two $f$ valley-orbitals per MUC and four $c$ bands per spin.
(For brevity, we call $f$ ``orbitals'' also for the VTHF model, in the rest of this paper.)
Its Hamiltonian is 
\begin{align}
& H_\VTHF = \sum_{\mb{R}} \frac{U_1}{2} N_\mb{R}^f (N_\mb{R}^f - 1)
+ \sum_{\mb{k} a a' \sigma} h^c_{a a'} (\mb{k}) c_{\mb{k}a\sigma}^\dagger c_{\mb{k}a'\sigma}
\nonumber \\
&
\quad + \sum_{\mb{k} a l \sigma} (h^{cf}_{al} (\mb{k}) c_{\mb{k}a\sigma}^\dagger f_{\mb{k} l \sigma} + \mr{h.c.})
- \mu N
+ H_{\nonloc} ,
\label{eq:H_VTHF}
\end{align}
where $f_{\mb{R}l\sigma}^\dagger$ creates a spin-$\sigma$ electron at the $f$ orbital of angular momentum $l = \eta (-1)^{\alpha-1} = \pm 1 \; (\mr{mod} \; 3)$ centered at position $\mb{R}$
and $f_{\mb{k}l\sigma}^\dagger$ is its Fourier transform.
(In single-site DMFT, the $C_{3z}$ rotation symmetry is promoted to the $\mr{U}(1)_\mr{orbital}$ symmetry~\cite{Wang2024:Molecular}.)
$c_{\mb{k}a\sigma}^\dagger$ creates a spin-$\sigma$ electron with momentum $\mb{k}$ in the $c$ band of index $a = 1, 2, 3, 4$.
$N_\mb{R}^f =  \sum_{l\sigma} f_{\mb{R}l\sigma}^\dagger f_{\mb{R}l\sigma}$
and $N$ count the $f$ electrons at $\mb{R}$ and all electrons, respectively, with respect to CNP.

The strength $U_1 = 57.95$ \meV of the $\mr{SU}(4)$ symmetric Hubbard interaction and the parameters for the nonlocal interactions $H_\nonloc$ remain the same as in the original THF model~\cite{Song2022}, while the $c$-band dispersion $h^c$ and the $c$-$f$ hybridization $h^{cf}$ may depend on the choice of the valley-orbital order~\cite{EndMatter}.
We consider the flat band limit~\cite{Song2022, Zhou2024} in which $h^c$ describes massless Dirac dispersion and thus $H_\VTHF$ is indistinguishable between the VP and KIVC cases, putting aside different basis transformations $(\alpha, \eta) \to l$.
We tune the chemical potential $\mu$ to study a range of $\nu = \la N \ra / n_M = \nu^f + \nu^c \in [0.8, 3.2]$ and $\nu^f = \la \sum_\mb{R} N^f_\mb{R} \ra / n_M$, where $n_M$ is the number of MUCs.

\textit{Effective Hund couplings from EPC.---}%
Similarly as in fullerides~\cite{Capone2001,Capone2002,Capone2004,Capone2009,Nomura2015},
the dynamical Jahn--Teller effect introduces spin anti-Hund ($J_S$) and orbital-angular-momentum Hund ($J_L$) couplings~\cite{Dodaro2018, Angeli2019, Wang2024:Molecular, Shi2024, Wang2024} to the VTHF model,
\begin{equation}
\begin{aligned}
H_\EPC &= \sum_\mb{R} \big[
2J_S \mb{S}_{\mb{R},+1} \cdot \mb{S}_{\mb{R},-1} - J_L (L_{\mb{R}}^2 - N_\mb{R}^f)
\\
& \phantom{=} + \tilde{U} N_\mb{R}^f (N_\mb{R}^f - 1) /2 \big],
\end{aligned}\label{eq:H_EPC}
\end{equation}
where $\mb{S}_{\mb{R}l}$ is the $f$-electron spin operator for orbital $l$ at $\mb{R}$ and $L_\mb{R} = \sum_{l\sigma} l f_{\mb{R}l\sigma}^\dagger f_{\mb{R}l\sigma}$ measures $f$-electron orbital angular momentum at $\mb{R}$~\cite{Wang2024, Wang2024:Molecular}.
The parameters of $H_\EPC$ depend on the order:
$(J_S, J_L, \tilde{U}) = (0.610, 0.069, 0.648)$, $(0.062, 0.605, 0.511)$, $(0.672, 0.084, -0.530)$ \meV for the VP, KIVC, and TIVC orders, respectively~\cite{EndMatter}.

To study phases at $\nu > 0$, we obtain symmetric (i.e., paramagnetic) normal-state solutions of $H_\VTHF$, with and without $H_\EPC$, for the VP, KIVC, and TIVC orders, by treating the correlations from Hubbard $U_1$ and $H_{\EPC}$ dynamically with DMFT+NRG and those from $H_{\nonloc}$ statically with Hartree mean-field approximation~\cite{Zhou2024, Youn2025, EndMatter}.
Since $H_\VTHF$ is identical and $H_\EPC$ differs between VP and KIVC, there are five distinct scenarios.
The results for $\nu < 0$ are obtained by particle-hole transformation~\cite{Bernevig2021:TBG3,Song2022,Zhou2024} $(\nu,\omega) \leftrightarrow (-\nu,-\omega)$, where $\omega$ is frequency.

\begin{figure}
\centerline{\includegraphics[width=\linewidth]{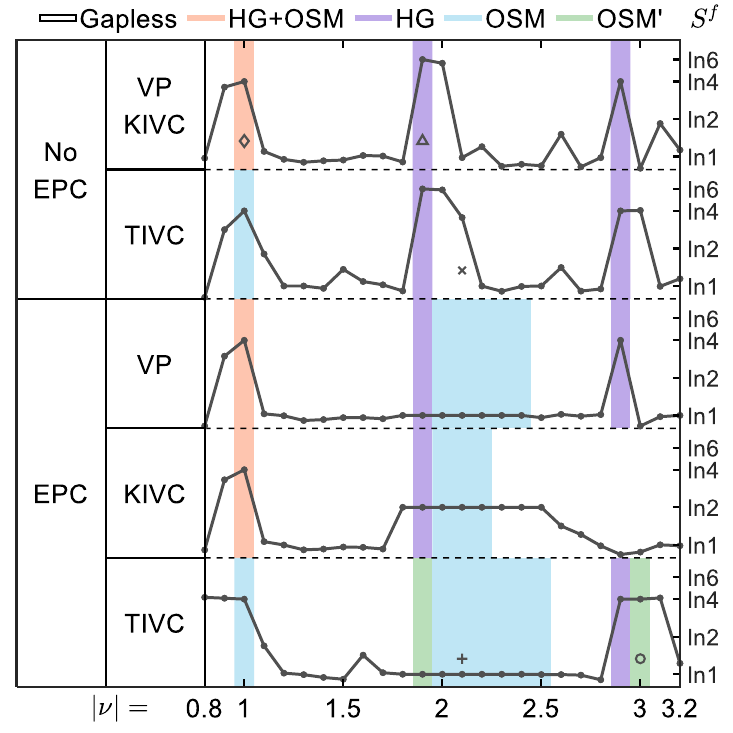}}
\caption{
Phase diagram of the VTHF model at $T = 10^{-6}$ \meV as a function of filling $\nu$, for different valley-orbital orders without and with EPC. In white regions, the spectral function is gapless.
Shades indicate phases with hard gaps (HG+OSM and HG) such that $A^f_{\tot}(\omega = 0) < 5\times10^{-4} /\meV$, $A^c_{\MBZ}(0) < 10^{-5} /\meV$ and phases with pseudogaps (OSM and \OSMp) such that $A^f_{\tot}(0) < 0.016 /\meV$, $A^c_{\MBZ}(0) < 0.025/\meV$.
Here $A^f_{\tot}(\omega) = \sum_{l\sigma} A^f_{l\sigma} (\omega)$ and $A^c_{\MBZ}(\omega) = n_M^{-1/2}\sum_{\mathbf{k}\in\MBZ}\sum_{a\sigma}A^c_{a\sigma}(\mathbf{k},\omega)$ are local $f$- and $c$-electron spectral functions, respectively, and $\mathrm{1MBZ}$ means the first moir\'{e} Brillouin zone.
Solid lines represent the $f$-electron contribution to the entropy per MUC, $S^f$, computed as the impurity entropy of the self-consistent impurity model~\cite{Costi1994} within DMFT.
Spectral properties for five states specified by markers are shown in \Fig{fig:phase_detail}.
}
\label{fig:phase_diagram}
\end{figure}

\begin{figure*}
\centerline{\includegraphics[width=\linewidth]{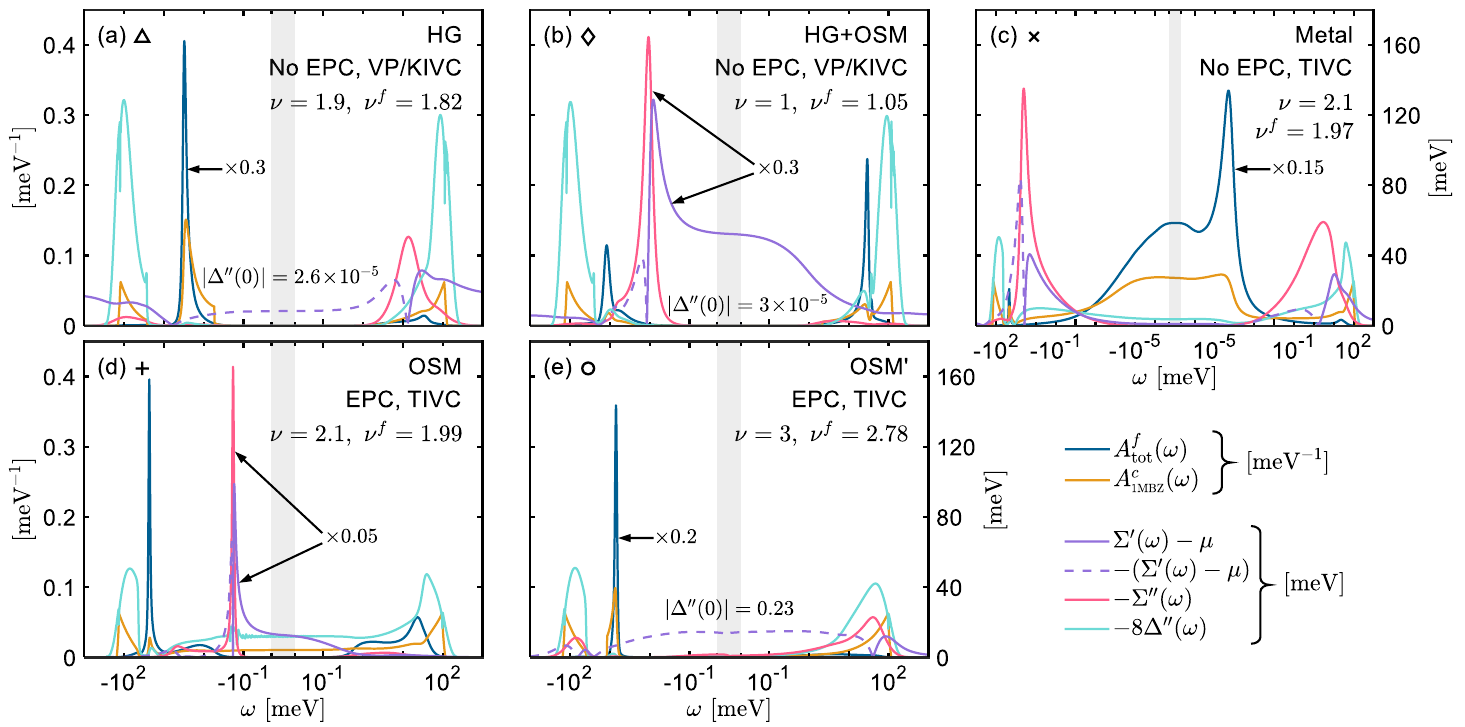}}
\caption{
Spectral properties of the VTHF model for parameters denoted by markers in the phase diagram (\Fig{fig:phase_diagram}).
Local spectral functions $A^f_{\tot}(\omega)$ (blue) and $A^c_{\MBZ}(\omega)$ (orange) are in units of $\meV^{-1}$ (left ordinate), while the self-energy shifted by the chemical potential, $\Sigma(\omega) - \mu$ (purple for the real part, magenta for the imaginary part), and the hybridization function $\Delta (\omega)$ (cyan for the imaginary part) are in \meV (right ordinate).
Purple dashed lines represent the sign-flipped negative part of $\Sigma'(\omega)-\mu$.
The abscissae are on a symmetric-log scale, with a narrow region (gray shade) on a linear scale that interfaces the sides of different signs.
In panels (a), (c), and (e) [(b) and (d)], the original data of $A^f_{\tot}(\omega)$ [$\Sigma(\omega) - \mu$] are downscaled to the plotted data by multiplying the factors specified next to the curves.
}
\label{fig:phase_detail}
\end{figure*}

\textit{Gap classification.---}%
We explain how we classify the $f$-electron spectral gap.
Within the DMFT description of a heavy-fermion type model (including the THF and VTHF models), the local $f$-electron Green's function is given by $\Gf(\omega)=[\omega +\mu-\Sigma(\omega)-\Delta(\omega)]^{-1}$, where $\Sigma(\omega)=\Sigma'(\omega)+\mathrm{i}\Sigma''(\omega)$ is the self-energy including the static Hartree energy from nonlocal interactions, and $\Delta(\omega)=\Delta'(\omega)+\mathrm{i}\Delta''(\omega)$ is the hybridization function between the $f$ orbital and the rest of the system.
Here we omit $l, \sigma$ indices since we focus on symmetric solutions.

The local $f$-electron spectral function,
\begin{equation}
 \Af(\omega)=\dfrac{-\frac{1}{\pi}(\Sigma''(\omega)+\Delta''(\omega))}{(\omega\!+\!\mu\!-\!\Sigma'(\omega)\!-\!\Delta'(\omega))^2 + (\Sigma''(\omega)\!+\!\Delta''(\omega))^2},
\label{eq:A^f}
\end{equation} 
opens a gap with a small $A^f (\omega = 0)$, when the numerator is small, or the denominator is large on the RHS.
(i) For the former, both $\Sigma''(0)$ and $\Delta''(0)$ must vanish.
That is, $\Delta''(\omega)$ has a gap at $\omega = 0$, which we call \textit{hybridization gap} (HG) [cf.~\Fig{fig:phase_detail}~(a)].
The latter case holds when $|\Sigma(0) - \mu|$ or $|\Delta(0)|$ is large.
(ii) In the \textit{OSM} phase~\cite{Sun2005,Tanaskovic2011,DeLeo2008:PRL,DeLeo2008:PRB,Gleis2024}, a self-energy pole ($\Sigma$ pole) appears as a sharp peak in $-\Sigma'' (\omega)$, accompanied by (via the Kramers--Kronig relations) a pair of negative and positive peaks in $\Sigma' (\omega)$ next to it [cf.~\Fig{fig:phase_detail}~(b,d)].
When the $\Sigma$ pole is at or near $\omega = 0$, $|\Sigma' (0)-\mu|$ or $-\Sigma''(0)$ becomes huge, respectively, leading to a small $A^f (\omega = 0)$.
(iii) We find a situation where $|\Sigma'(0)-\mu|$ is large and $\Sigma$ strongly depends on $\omega$, yet there is no self-energy pole [cf.~\Fig{fig:phase_detail}~(e)];
we coin it \textit{\OSMp} based on its similarity to and difference from OSM.
(iv) Lastly, a \textit{Kondo insulator} is characterized by a pole in $\Delta$ near or at $\omega = 0$~\cite{Georges1996,Pruschke2000,Bulla2008}.
We emphasize that these gap-opening mechanisms may coexist since the conditions are not exclusive, except for OSM vs \OSMp.

\textit{Phase diagram.---}%
We apply the above criteria to identify the phases of the VTHF model, as shown in Fig.~\ref{fig:phase_diagram}.
We focus on $T = 10^{-6}$ \meV, much lower than experimentally accessible temperatures.
When EPC is off, we find hard-gap phases at $|\nu| = 1, 1.9, 2.9$, except that the TIVC case at $|\nu| = 1$ has a pseudogap.
They feature large $f$-electron contributions to the entropy per MUC, $S^f = \ln 4$ at $|\nu| = 1, 2.9$ and $S^f = \ln 6$ at $|\nu| = 1.9$, which reflect the four- and six-fold degeneracies of the $\mr{SU}(4)$ Hubbard term for one (or three, equivalently) and two particles, respectively.
The HG phases at $|\nu| = 1.9, 2.9$ are characterized by a gap of $\Delta''(\omega)$ at $\omega = 0$ [\Fig{fig:phase_detail}(a)].
The HG+OSM phase for VP and KIVC at $|\nu| = 1$ exhibits a $\Sigma$ pole at $\omega \simeq -1 \, \meV$ induced by the Hubbard interaction, on top of the HG [\Fig{fig:phase_detail}(b)].
A similar $\Sigma$ pole makes the TIVC case at $|\nu| = 1$ an OSM phase.

Turning on EPC strongly modifies the system at $1.9 \lesssim |\nu| \lesssim 2.5$, where $|\nu^f|$ stays close to 2 due to the strong Hubbard repulsion~\cite{Hu2023, Rai2024}.
For the VP and TIVC orders, $S^f$ vanishes since the spin anti-Hund coupling in $H_\EPC$ with $J_S \gg J_L$ favors the singlet state $(l, s) = (0, 0)$ of two $f$ electrons (holes) for $\nu^f = 2$ $(-2)$, which is similar to a crystalline-electric-field (CEF) singlet~\cite{Yotsuhashi2002}.
Here $l$ and $s$ are the quantum numbers associated with the $\mr{U}(1)_\mr{orbital}$ and $\mr{SU}(2)_\mr{spin}$ symmetries, respectively. 
On the other hand, for the KIVC order, the orbital-angular-momentum Hund coupling in $H_\EPC$ with $J_S \ll J_L$ favors the orbital doublet $(l, s) = (\pm 2, 0)$, resulting in $S^f = \ln 2$.

Importantly, the OSM phase at $2 \leq |\nu| \leq \nu_{\mr{cr}}$ ($\nu_{\mr{cr}} = 2.4, 2.2, 2.5$ for VP, KIVC, and TIVC, respectively) is consistent with experimental observations of robust pseudogap in the AA-stacking region at a similar range of $\nu$~\cite{Oh2021,Nuckolls2023}.
Without EPC and further symmetry breaking (beyond that imposed by the downfolding), the system at $2 \leq |\nu| \leq 2.5$ features $f$- and $c$-electron spectral functions peaking around $\omega = 0$ [cf.~\Fig{fig:phase_detail}(c)]~\cite{Zhou2024}.
Meanwhile, EPC introduces a $\Sigma$ pole near $\omega = 0$ that opens a gap in the spectral function [\Fig{fig:phase_detail}(d)].
The gap formation can be explained by the atomic charge gap increased by $H_\EPC$~\cite{EndMatter}, reminiscent of the Janus effect in Hund metals~\cite{Werner2009,deMedici2011:PRB,deMedici2011:PRL,Georges2013,Stadler2019}:
In $t_{2g}$ and $e_g$ systems at half filling, the Hund coupling increases the atomic gap and decreases the critical interaction strength of the Mott transition, favoring an insulating phase;
away from half filling, the same coupling decreases the atomic gap, favoring metallicity.
We emphasize that the valley-orbital ordering is also crucial, as it shifts ``half'' filling from $\nu^f = 0$ for the original THF model to $|\nu^f| = 2$ for the VTHF model.

By contrast, the phases at $|\nu| \simeq 1, 3$ are overall less affected by EPC, while the gap disappears in the KIVC case at $|\nu| = 2.9$ and the \OSMp phases arise in the TIVC case at $|\nu| = 1.9, 3$.
Despite the absence of $\Sigma$ poles, the \OSMp phase is strongly correlated; for example, $\Sigma (0)$ clearly deviates from the Hartree value $\Sigma (\pm\infty)$ in \Fig{fig:phase_detail}(e).

The gapless phase (white regions in \Fig{fig:phase_diagram}) is peculiar since $|S^f|$ is small but nonzero.
By analyzing the energy flow diagram and the scaling of $\Sigma''(\omega)$, we identify them as incoherent metals, not Fermi liquids.
$S^f>0$ implies that $f$ electrons are not fully screened since the Fermi-liquid coherence scale is exponentially low ($\lesssim T = 10^{-6} \, \meV$) due to the small hybridization strength $|\Delta'' (0)|$ originating from the small density of states of the Dirac bands near the Fermi level.
Such small $|\Delta''|$ is also the reason for the absence of a Kondo insulator phase in our phase diagram.
On the other hand, $S^f < 0$ comes from the strong $\omega$-dependence of $\Delta''$~\cite{Zhuravlev2009, Zhuravlev2011, Mastrogiuseppe2014, Wu2022, Shankar2023}.
See Ref.~\cite{Youn2025} for details.

\begin{figure}
\centerline{\includegraphics[width=\linewidth]{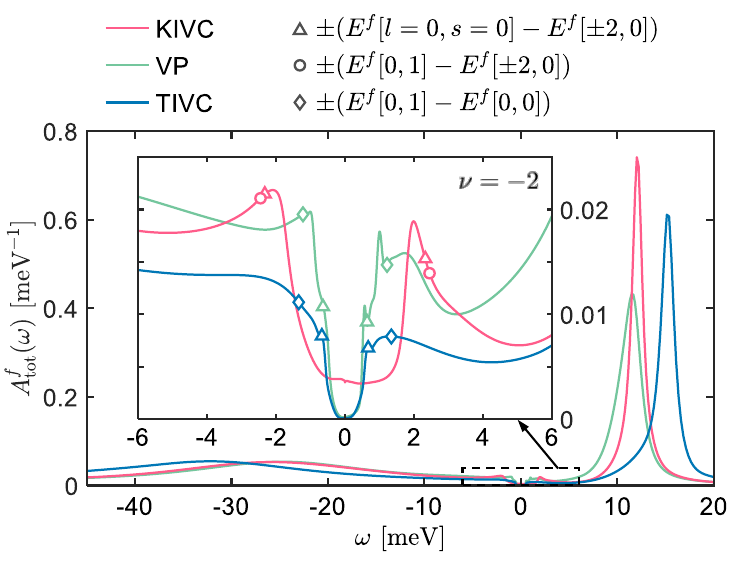}}
\caption{
Local $f$-electron spectral function at $\nu = -2$ for different orders with EPC, which corresponds to the local density of states in the AA-stacking region.
The inset zooms into the low-frequency region, $|\omega|<6$ \meV, which shows pseudogaps.
Markers indicate the differences between the lowest-lying energy eigenvalues $E^f$ of the local $f$-electron Hamiltonian.
$l$ and $s$ are orbital and spin quantum numbers, respectively.
}
\label{fig:Vshape}
\end{figure}

\begin{figure}
\centerline{\includegraphics[width=\linewidth]{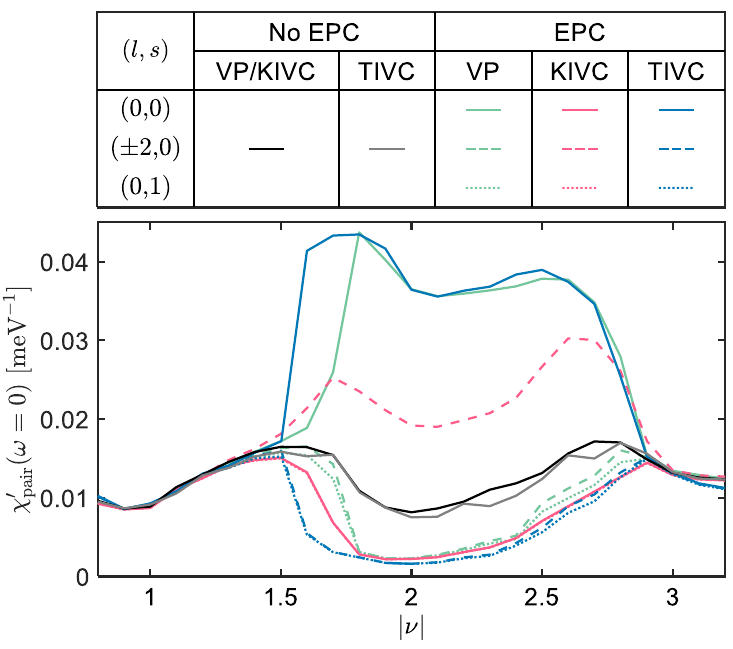}}
\caption{
Static pairing susceptibilities for $f$-electron pairs with different symmetries, denoted by the quantum numbers $(l, s)$ of the pair creation operators.
When EPC is off, the curves for different $(l,s)$ are identical due to the $\mr{SU}(4)$ symmetry. 
Solid and dotted magenta lines lie on top of each other.
}
\label{fig:pairing_susc}
\end{figure}

\textit{Pseudogap and pairing susceptibilities.---}%
The $f$-electron pseudogap with side peaks at its edges appears throughout the OSM phase at $2 \leq |\nu| \leq \nu_\mr{cr}$; see \Fig{fig:Vshape} for the case of $\nu = -2$.
The positions of the side peaks align well with spin, orbital, or spin-orbital excitation energy scales~\cite{Youn2025}, which are also given by the splitting of the two-hole (cf.~$\nu, \nu^f < 0$) eigenvalues of the local $f$-electron Hamiltonian, according to the effective Hund couplings in $H_\EPC$.
In the VP and TIVC cases, the two-hole ground state is the singlet $(l, s) = (0, 0)$.
The other two-hole eigenstates are the orbital doublet $(\pm 2, 0)$ and the spin triplet $(0, 1)$.
Thus the energy differences $E^f [l = \pm 2, s = 0] - E^f [0,0]$ and $E^f [0, 1] - E^f [0,0]$ mean, respectively, the local orbital and spin excitation energies, around which the side peaks are located.
A similar argument holds for the KIVC case, where $(\pm 2, 0)$ is the ground-state doublet, while $(0,0)$ and $(0,1)$ are excited states.

Figure \ref{fig:pairing_susc} displays the static pairing susceptibilities $\chi'_{\mr{pair}} (\omega=0)$ for different symmetries $(l, s)$ of an $f$-electron pair.
For $\nu > 0$, we compute $\chi'_{\mr{pair}} (0)$ as the real part of $\chi_{\mr{pair}} (\omega) = -\mi \int_0^\infty \mr{d}t \, \mr{e}^{\mi (\omega+\mi 0^+)t}\la[P_{ls} (t),P_{ls}^\dagger (0)]\ra$ in the active space, where $P_{ls}^\dagger$ creates two $f$ electrons such that $P_{ls}^\dagger \ket{\CNP}$ has quantum numbers $(l,s)$, and $\ket{\CNP}$ is the vacuum of the active space.
Notably, EPC enhances $\chi'_{\mr{pair}} (0)$ at $1.6 \lesssim \nu \lesssim 2.8$ for a specific $(l,s)$ component for each valley-ordering, 
such that the action of $P_{ls}^\dagger$ to the local two-$f$-electron ground state leads to full $f$ orbitals.
Here $l = 0$ ($\pm 2$) pair creation implies $s$-($d$-)wave pairing~\cite{EndMatter}.
The case for $\nu < 0$ is similar, after particle-hole transformation.

\textit{Discussion.---}%
The OSM phase near $|\nu| = 2$ and the pseudogap with the side peaks, both induced by EPC, constitute the central result of this work.
It is another evidence of Hundness---strong correlations arising from (effective) Hund interactions---that the side peaks are associated with spin, orbital, or spin-orbital fluctuations, evocative of orbital resonance features in the spectral function of Hund metals~\cite{Stadler2015,Kugler2019,Stadler2019,Kugler2020}.
Since those fluctuations are charge-neutral, the side peaks cannot be captured by Hartree--Fock (HF) solutions such as the TIVC$\times$QSH (quantum spin Hall) order~\cite{Kwan2023, Wang2024, Shi2024}.

Among the three valley-orbital orders, we conclude that the TIVC result is most consistent with experimental observations at $|\nu| \simeq 2$.
$S^f = 0$ means that excess electrons (or holes) do not form moments and the underlying TIVC order is exposed in the whole system.
This result reconciles the STM imaging of TIVC order~\cite{Nuckolls2023} and the spin-singlet Mott-like feature~\cite{Cao2018correlated}.
Also, the width of the $f$-electron pseudogap is consistent with the STM measurements on the AA-stacking region~\cite{Oh2021}.

On the other hand, we find large spin-valley-orbital moments corresponding to $S^f = \ln 4$ for general scenarios at $|\nu| \simeq 1, 3$, except for the KIVC case with EPC at $|\nu| \simeq 3$.
Given that the $f$-electron dispersion is very flat, the RKKY interactions between these moments will be ferromagnetic, which can explain the observed symmetry-breaking states at $|\nu| = 1,3$ in experiments~\cite{Lu2019,grover_chern_2022,stepanov_competing_2021}.

The range of $1.6 \lesssim \nu \lesssim 2.8$ for enhanced pairing susceptibilities is consistent with the superconducting domes~\cite{Cao2018unconventional, Cao2020, Oh2021, cao_nematicity_2021,Nuckolls2023,tanaka_superfluid_2024}.
These pairs are local, so they may be a precursor to molecular pairing~\cite{Wang2024:Molecular}, among various proposals of pairing glue~\cite{Wu2018, Lian2019, Liu2024, Yu2022, Eslam2021, Blason2022, Christos2023, Islam2023}.
However, we cannot decisively fix the pairing symmetry of MATBG from our result since valley-orbital fluctuations, which we ignored by choosing the static order, may be important to stabilize pairing.
It will be worthwhile to obtain the solutions with broken $\mr{U}(1)_\mr{charge}$ symmetry, considering a breakdown of valley-orbital ordering and the role of strain, which is left to future studies.

\begin{acknowledgments}
We thank Andreas Gleis, Minsoo Kim, Fabian Kugler, Jong Yeon Lee, Mathias Pelz, Yi-Jie Wang, and Jan von Delft for fruitful discussions.
This work was supported by Samsung Electronics Co., Ltd.~(No.~IO220817-02066-01), the Global-LAMP Program of the National Research Foundation of Korea (NRF) grant funded by the Ministry of Education (No.~RS-2023-00301976), the NRF grants funded by the Korean government (MSIT) (No.~RS-2023-00214464, No.~RS-2023-00258359, No.~RS-2023-NR119931, No.~RS-2024-00442710), and the NRF grant funded by the Korean government (MEST) (No.~2019R1A6A1A10073437).
Z.-D.S.~and G.-D.Z.~were supported by National Natural Science Foundation of China (General Program No.~12274005), National Key Research and Development Program of China (No.~2021YFA1401903), and Innovation Program for Quantum Science and Technology (No.~2021ZD0302403).
\end{acknowledgments}

\bibliography{references} 
\clearpage
\onecolumngrid
\section{End Matter}
\twocolumngrid

\textit{Details of $H_\VTHF$.---}%
We explain the derivation of $H_\VTHF$ first for the VP case.
The $f$ electrons in the VP state at CNP have the same valley, say $-$ without loss of generality.
Then the active, excess $f$ electrons at $\nu>0$ are all from the $+$ valley, 
$(f_{\mb{R},+1,\sigma}, f_{\mb{R},-1,\sigma}) = (f_{\mb{R},1,+,\sigma}^\THF, f_{\mb{R},2,+,\sigma}^\THF$),
where $f_{\mb{R}l\sigma}$ and $f_{\mb{R}\alpha\eta\sigma}^\THF$ annihilate, respectively, an active $f$ electron from the VTHF model and an $f$ electron from the unprojected THF model.
($\alpha = 1,2$, $\eta = \pm$, $l = \pm 1$, $\sigma = \ua, \da$) Projecting the non-interacting part of the THF model Hamiltonian in the flat band limit~\cite{Song2022, Zhou2024} onto the active space, we get 
\begin{equation}
\begin{aligned}
h^c(\mb{k}) &=
\begin{pmatrix}
  0_{2\times2}  & v_{\star}(k_x \sigma_0+\mr{i}k_y\sigma_z)\\
   v_{\star}(k_x\sigma_0-\mr{i}k_y\sigma_z) & 0_{2\times2}
\end{pmatrix}
, \\
h^{cf}(\mb{k}) &= e^{-\lambda^2|\mathbf{k}|^2}[\gamma\sigma_{0}+v_{\star}'(k_x\sigma_x+k_y\sigma_y), \,
0_{2\times2}]^{\dagger},
\end{aligned}
\label{eq:nonint_H_VTHF}
\end{equation}
whose rows and columns are indexed by $a = 1,2,3,4$, except for the columns of $h^{cf}(\mb{k})$ indexed by $l = +1, -1$.
$\sigma_{0,x,y,z}$ are the Pauli matrices.
We adopt the parameters from Ref.~\cite{Song2022}: $v_{\star}=-4303$~meV$\cdot$\AA, $\lambda=0.414/\sin(\frac{\theta_m}{2})$\AA~with magic angle $\theta_m=1.05^\circ$, $\gamma=-24.75$~meV, and $v_{\star}'=1622$~meV$\cdot$\AA.
The interacting part of $H_\VTHF$ is obtained by projecting that of the THF model onto the active space, after treating the inactive $f$ and $c$ electrons as a static background. The resulting interactions are the $\mr{SU}(4)$ Hubbard $U_1$ interaction and $H_\nonloc$, comprising the $c$-$c$ density-density ($V$), $c$-$f$ density-density ($W_{a = 1,2,3,4}$), and $c$-$f$ exchange ($J$) interactions. (We neglect the nearest-neighbor $f$-$f$ interaction ($U_2$)~\cite{Rai2024}.)

The KIVC (TIVC) state, which has the order parameter $\la\sigma_y\tau_y\ra$ ($\la\sigma_x\tau_x\ra$), is obtained by applying a flat (chiral) $\mr{U}(4)$ rotation~\cite{Song2022} to the VP state.
Here $\sigma_{0,x,y,z}$ and $\tau_{0,x,y,z}$ are the Pauli matrices for orbital ($\alpha$) and valley ($\eta$), respectively.
We choose 
$f_{\mb{R},\pm1,\sigma} = \frac{1}{\sqrt{2}}(f_{\mb{R},1,\pm,\sigma}^\THF \mp f_{\mb{R},2,\mp,\sigma}^\THF)$ for KIVC
and
$f_{\mb{R},\pm1,\sigma} = \frac{1}{\sqrt{2}}(f_{\mb{R},1,\pm,\sigma}^\THF + f_{\mb{R},2,\mp,\sigma}^\THF)$ for TIVC~\cite{Song2022,Youn2025}.
The resulting $H_\VTHF$ for the KIVC or TIVC case has $h^c$, $h^{cf}$, Hubbard $U_1$, and $H_{\nonloc}$ identical to those for the VP case, except that the TIVC takes $v_{\star}'=0$~\cite{Youn2025}.

\textit{Details of $H_\EPC$.---}%
The effective local $f$-electron interaction arises in the full THF model, by integrating out all ($\Gamma$ and $K$, acoustic and optical) phonon degrees of freedom~\cite{Wang2024}.
Projecting the interaction term [cf.~Eqs.~(72)--(76) of Ref.~\cite{Wang2024}] onto the active space~\cite{Wang2024, Youn2025}, we get
\begin{widetext}
\begin{equation}
\begin{aligned}
H_\EPC &= 
\frac{1}{2} \sum_{\mathbf{R}\sigma \sigma'} 
\begin{pmatrix}
-J_a' &\! 0 &\! 0 &\! 0 \\
0 &\! J_c' &\! -J_b' &\! 0 \\
0 &\! -J_b' &\! J_c' &\! 0 \\
0 &\! 0 &\! 0 &\! -J_a' \\
\end{pmatrix}_{m'l', ml}
f^{\dagger}_{\mathbf{R} l' \sigma} f^{\dagger}_{\mathbf{R} m' \sigma'} f_{\mathbf{R} m \sigma'} f_{\mathbf{R} l \sigma}
\\
&=
\sum_{\mathbf{R}}
\Bigg[ -J'_a \sum_{l = \pm 1}
\frac{N^f_{\mathbf{R}l}(N^f_{\mathbf{R}l}-1)}{2}
+J_b'
\bigg(\frac{N^f_{\mathbf{R},1}N^f_{\mathbf{R},-1}}{2}
+2\mathbf{S}_{\mathbf{R},1} \cdot \mathbf{S}_{\mathbf{R},-1}\bigg) 
+J_c'N^f_{\mathbf{R},1}N^f_{\mathbf{R},-1}\Bigg],
\end{aligned}
\label{eq:H_EPC_long}
\end{equation}
where the rows and columns of the $4 \times 4$ matrix in the first line are indexed by 
$(m^{(\prime)}, l^{(\prime)}) = (1,1), (1,-1), (-1,1), (-1,-1)$,
and $N^f_{\mb{R}l} = \sum_{\sigma} f_{\mb{R}l\sigma}^\dagger f_{\mb{R}l\sigma}$.
The strengths of $J'_{a,b,c}$ depend on the valley-orbital order,
\begin{equation}
(J'_a, J'_b, J'_c) =
\left\{
\begin{matrix}
\big( \frac{1}{2}J_a+K, & J_b+K, & \frac{1}{2}J_a \big), & \text{(VP)}, \\
\big( \frac{\lambda_{\RG}}{2}J_d+K, & \frac{\lambda_{\RG}}{2}J_e+\frac{3}{2}K, & \frac{\lambda_{\RG}}{2}J_d \big), & \text{(KIVC)}, \\
\big( \frac{\lambda_{\RG}}{2}J_d+K, & J_b+\frac{\lambda_{\RG}}{2}J_e+\frac{5}{2}K, & -\frac{\lambda_{\RG}}{2}J_d-K \big), & \text{(TIVC)}, \\
\end{matrix}
\right.
\end{equation}
\end{widetext}
where $(J_a, J_b,J_d, J_e)=(0.96,1.60,1.30,1.19) \, \meV$ and $\lambda_\RG \in [1, 3.2]$ is the renormalization effect from electrons and $K$-phonon states~\cite{Wang2024, Basko2008}.
$K=-0.11\times10^{-3}U_0$ depends on the on-site Hubbard repulsion $U_0 \in [3, 9] \, \text{eV}$ of carbon atom~\cite{Wehling2011, Schuler2013} [cf.~Eqs.~(84)--(87) of Ref.~\cite{Wang2024}]. We use $\lambda_{\RG}=2.6$ and $U_0=9 \, \text{eV}$ in our calculation. Rephrasing \Eq{eq:H_EPC_long}, we get $J_S =J_b'$, $J_L =\frac{1}{4}J_a'+\frac{1}{8}J_b'+\frac{1}{4}J_c'$, and $\tilde{U} = -\frac{1}{2}J_a'+\frac{1}{4}J_b'+\frac{1}{2}J_c'$ for Eq.~(\ref{eq:H_EPC}) in the main text.

\textit{DMFT details.---}%
Our DMFT self-consistency loop is nested, having outer and inner loops.
The outer loop involves the $\mb{k}$-space integration of correlators and the NRG impurity solver step.
After applying Hartree mean-field decoupling to $H_\nonloc$, our effective impurity model has two active $f$ orbitals as impurity and the corresponding two bath channels.
After the NRG obtains $\Sigma (\omega)$, we run the inner, charge self-consistency loop, where $\mu$ is determined so that $\nu^f$ and $\nu^c$ are self-consistent in consideration of the Hartree mean-fields and satisfy $\nu = \nu^f + \nu^c$, for a given target $\nu$.

In the THF and VTHF models, the $f$ orbitals form a lattice, but the $c$ bands form a continuum.
That is, the $\mb{k}$-space of the $c$ bands is unbounded.
When we evaluate $G^f (\mb{k} \in \mr{1MBZ},\omega)$ in the outer loop, we include the contribution from $[G^{c}(\mb{k}\!+\!\mb{G},\omega)]^{-1}$ for $\mb{G} = 0$ (1MBZ) and also $\mb{G}$'s pointing to six reciprocal primitive cells adjacent to the 1MBZ, which is necessary to capture the full $\omega$ dependence of the hybridization function $\Delta(\omega)$.
If only $\mb{G} = 0$ is considered, then the the high-frequency part of $\Delta(\omega)$ gets truncated, which leads to artifacts in the Kondo temperature, filling, etc.

\textit{NRG details.---}%
We use the logarithmic discretization parameter $\Lambda = 4$ and perform $z$ averaging over $n_z = 2$ discretization grids~\cite{Zitko2009}.
We exploit $\mr{U}(1)_\mr{charge} \times \mr{SU}(4)_\mr{spin, orbital}$ [$\mr{U}(1)_\mr{charge} \times \mr{U}(1)_\mr{orbital} \times \mr{SU}(2)_\mr{spin}$] symmetry when EPC is off (on), by using the QSpace tensor library~\cite{Weichselbaum2012:QSpace1, Weichselbaum2020:QSpace2, Weichselbaum2024:QSpace3}.
In iterative diagonalization, up to $N_\mr{keep} = 3{,}500$ $(5{,}000)$ symmetry multiplets are kept when EPC is off (on).
Discrete spectral data of correlation functions are obtained using the full-density-matrix NRG~\cite{Weichselbaum2007,Weichselbaum2012:NRG} and then adaptively broadened~\cite{Lee2016:MuNRG1}.
We use the symmetric improved estimator of the self-energy~\cite{Kugler2022} that greatly ameliorates the accuracy for given $N_\mr{keep}$.

\textit{Atomic charge gap.---}%
Table~\ref{tab:f_states} lists the eigenstates and eigenvalues of the local $f$-electron interactions (Hubbard $U_1$ and $H_\EPC$).
The atomic charge gap at integer filling $\nu^f$ is given by $\Delta_\mr{at} = E^f_{\min} [n^f = \nu^f + 1] + E^f_{\min} [n^f = \nu^f - 1] - 2 E^f_{\min} [n^f = \nu^f]$, where $E^f_{\min} [n^f]$ means the lowest energy for a given $n^f$.
Hence we get
\begin{equation}
\begin{aligned}
\left. \Delta_\mr{at} \right|_{n^f = 1, 3} &=
\left\{
\begin{matrix}
U_1 - \frac{3}{2} J_S + 2J_L + \tilde{U}, & \text{(VP, TIVC)}, \\
U_1 - 2 J_L + \tilde{U}, & \text{(KIVC)}, \\
\end{matrix}
\right.
\\
\left. \Delta_\mr{at} \right|_{n^f = 2} &=
\left\{
\begin{matrix}
U_1 + 3 J_S - 2J_L + \tilde{U}, & \text{(VP, TIVC)}, \\
U_1 + 6 J_L + \tilde{U}, & \text{(KIVC)}. \\
\end{matrix}
\right.
\end{aligned}
\end{equation}
We see $\left. \Delta_\mr{at} \right|_{n^f = 1, 3} < U_1$ and $\left. \Delta_\mr{at} \right|_{n^f = 2} > U_1$, a manifestation of the Janus effect of Hundness.

\textit{Pairing symmetries.---}%
We discuss the pairing symmetries associated with the favored pairing fluctuations by EPC in different orders.
The unprojected original THF model preserves $D_6$ point group, whose generators $g\in\{C_{3z},C_{2z},C_{2x}\}$ are represented in the unprojected ($f^{\THF}_{\alpha\eta\sigma}$) basis (here we omit index $\mb{R}$ for brevity) as $C_{3z}=e^{\mr{i}\frac{2\pi}{3}\sigma_z\tau_z},C_{2z}=\sigma_x\tau_x,C_{2x}=\sigma_x\tau_0$.

The TIVC order with order parameter $\la\sigma_x\tau_x\ra$ preserves the $D_6$ point group, with the symmetry operators on the projected ($f_{l\sigma}$) basis given by $C_{3z}=e^{\mr{i}\frac{2\pi}{3}\tilde{\sigma}_z},~C_{2z}=\tilde{\sigma}_0$, and $C_{2x}=\tilde{\sigma}_x$. Here, $\tilde{\sigma}_{0,x,y,z}$ are the Pauli matrices for the valley-orbital ($l$) of the active $f$ electrons. The EPC-favored $(l,s)=(0,0)$ pairing forms the $A_1$ representation, which equals (up to a normalization factor) to the
projected $s$-wave pairing of the unprojected model~\cite{Wang2024:Molecular}. While the $d$-wave state is the local two-particle ground state in the unprojected model under the chosen parameters (Fig.~6 of Ref.~\cite{Wang2024}), the projection omits off-diagonal components between the active and inactive bases and makes the $s$-wave state the ground state.
From Fig.~\ref{fig:pairing_susc}, we observe that pairing susceptibility is enhanced for channels associated with the local two-particle ground state. Therefore, incorporating the inactive basis or considering valley-orbital fluctuations may change the favored pairing channel.

The KIVC order with order parameter $\la\sigma_y\tau_y\ra$ breaks the $C_{2x}$ symmetry but preserves a projective $D_6$ (dubbed as $D_6'$) group generated by $C_{3z}=e^{\mr{i}\frac{2\pi}{3}\tilde{\sigma}_z},~C_{2z}=-\tilde{\sigma}_z$, and $C_{2x}'= \tilde{\sigma}_x$, where $C_{2x}'$ represents $C_{2x}$ followed by a valley $\pi$-rotation $\tau_z$. The EPC-favored $(\pm2,0)$ pairings form the $E_2$ representations of $D_6'$, corresponding (up to a normalization factor) to the projected $d$-wave $E_2$ pairing of the unprojected model.

This analysis can also be applied to the VP state, which breaks the $C_{2z}$ symmetry while respecting a projective $D_6$ group. In this case, the EPC-favored $(0,0)$ pairing is obtained by projecting the valley-polarized $A_1$ $s$-wave pairing of the unprojected model~\cite{Youn2025}.
\onecolumngrid

\begin{table}[h!]
\centering
\renewcommand{\arraystretch}{1.5}
\setlength{\tabcolsep}{5pt}      
\begin{tabular}{c|c|c|c|c|c}
\hline \hline
$n^f$ & $l$ & $s$ & Degeneracy & {$f$-electron eigenstates} & $E^f$ \\ \hline
0 & 0 & 0 & 1 & $\ket{\CNP}$ & $0$ \\ \cline{1-6}
1 & $\pm1$ & $\frac{1}{2}$ & 4 & $\{f^{\dagger}_{1\uparrow} ,~f^{\dagger}_{1\downarrow},~f^{\dagger}_{-1\uparrow},~f^{\dagger}_{-1\downarrow}\}\ket{\CNP}$ & $0$ \\ \cline{1-6}
\multirow{3}{*}{2} & $\pm2$ & 0 & 2 & $\{f^{\dagger}_{1\uparrow}f^{\dagger}_{1\downarrow},~f^{\dagger}_{-1\uparrow}f^{\dagger}_{-1\downarrow}\}\ket{\CNP}$ & \color[rgb]{1,0,0}{$U_1-2J_L+\tilde{U}$} \\ \cline{2-6}
  & \multirow{2}{*}{0} & 0 & 1 & $\frac{1}{\sqrt{2}}(f^{\dagger}_{1\uparrow}f^{\dagger}_{-1\downarrow}-f^{\dagger}_{1\downarrow}f^{\dagger}_{-1\uparrow})\ket{\CNP}$ & \color[rgb]{0,0,1}{$U_1-\frac{3}{2}J_S+2J_L+\tilde{U}$} \\  \cline{3-6}
  &  & 1 & 3 & $\{\frac{1}{\sqrt{2}}(f^{\dagger}_{1\uparrow}f^{\dagger}_{-1\downarrow}+f^{\dagger}_{1\downarrow}f^{\dagger}_{-1\uparrow}),~f^{\dagger}_{1\uparrow}f^{\dagger}_{-1\uparrow},~f^{\dagger}_{1\downarrow}f^{\dagger}_{-1\downarrow}\}\ket{\CNP}$ & $U_1+\frac{1}{2}J_S+2J_L+\tilde{U}$\\ \hline
3 & $\pm1$ & $\frac{1}{2}$ & 4 & $\{
f^{\dagger}_{1\uparrow}f^{\dagger}_{1\downarrow}f^{\dagger}_{-1\downarrow},~f^{\dagger}_{1\uparrow}f^{\dagger}_{1\downarrow}f^{\dagger}_{-1\uparrow},~
f^{\dagger}_{1\downarrow}f^{\dagger}_{-1\uparrow}f^{\dagger}_{-1\downarrow},~f^{\dagger}_{1\uparrow}f^{\dagger}_{-1\uparrow}f^{\dagger}_{-1\downarrow}
\}\ket{\CNP}$ & $3U_1+2J_L+3\tilde{U}$ \\ \hline
4 & 0 & 0 & 1 & $f^{\dagger}_{1\uparrow}f^{\dagger}_{1\downarrow}f^{\dagger}_{-1\uparrow}f^{\dagger}_{-1\downarrow}\ket{\CNP}$ & $6U_1+4J_L+6\tilde{U}$ \\ \hline \hline
\end{tabular}
\caption{%
The atomic energy eigenvalues and eigenstates of active $f$ electrons according to local interactions (Hubbard $U_1$ and $H_{\EPC}$).
$n^f$ is quantum number associated with the $\mr{U}(1)_\mr{charge}$ symmetry.
We omit the index $\mb{R}$ for the brevity.
The atomic ground-state energy for the KIVC (VP or TIVC) case with EPC is color-coded in red (blue). \ket{\mr{CNP}} denotes the inactive $f$-electron background at CNP.}
\label{tab:f_states}
\end{table}

\end{document}